\documentclass[aps,twocolumn,english,superscriptaddress,citeautoscript,showkeys,preprintnumbers,amsmath,amssymb,floatfix,footinbib,prl]{revtex4-2}
\usepackage{graphicx}
\usepackage{xcolor}
\usepackage{soul}
\usepackage{amssymb,amsmath}
\usepackage{hyperref}
\hypersetup{colorlinks=true,linkcolor=black,citecolor=black,urlcolor=black}
\usepackage[utf8]{inputenc}
\usepackage[english]{babel}
\usepackage{txfonts}\rm
\usepackage{booktabs}
\usepackage{siunitx}
\DeclareSIUnit\angstrom{\protect \text {\AA}}

\newcommand{\tc}{$T_\text{c}$}
\newcommand{\ef}{\varepsilon_{\mathrm{F}}}
\newcommand{\gap}{$\Delta_{n\textbf{k}}$}

\newcommand{\ML}{UF}
\newcommand{\AD}{mMc}

\def\bq{{\bf q} }
\def\bk{{\bf k} }

\def\cp{{\mu}}   
\def\cs{{\mu^*}} 

\begin{document}

\title{Full-bandwidth anisotropic Migdal-Eliashberg theory and its application to superhydrides}

\author{Roman Lucrezi}
\thanks{These authors contributed equally}
\affiliation{Institute of Theoretical and Computational Physics, Graz University of Technology, NAWI Graz, 8010, Graz, Austria}

\author{Pedro P. Ferreira}
\thanks{These authors contributed equally}
\affiliation{Institute of Theoretical and Computational Physics, Graz University of Technology, NAWI Graz, 8010, Graz, Austria}
\affiliation{Universidade de S\~ao Paulo, Escola de Engenharia de Lorena, DEMAR, Lorena, Brazil}

\author{Samad Hajinazar}
\affiliation{Department of Physics, Applied Physics, and Astronomy, Binghamton University-SUNY, Binghamton, New York 13902, USA}
\affiliation{Present address: Department of Chemistry, University at Buffalo, Buffalo, NY 14260, USA}

\author{\\Hitoshi Mori}
\affiliation{Department of Physics, Applied Physics, and Astronomy, Binghamton University-SUNY, Binghamton, New York 13902, USA}

\author{Hari Paudyal}
\affiliation{Department of Physics, Applied Physics, and Astronomy, Binghamton University-SUNY, Binghamton, New York 13902, USA}

\author{Elena R. Margine}
\email[Corresponding author: ]{rmargine@binghamton.edu}
\affiliation{Department of Physics, Applied Physics, and Astronomy, Binghamton University-SUNY, Binghamton, New York 13902, USA}

\author{Christoph Heil}
\email[Corresponding author: ]{christoph.heil@tugraz.at}
\affiliation{Institute of Theoretical and Computational Physics, Graz University of Technology, NAWI Graz, 8010, Graz, Austria}

\begin{abstract}
\textbf{Abstract:} Migdal-Eliashberg theory is one of the state-of-the-art methods for describing conventional superconductors from first principles. However, widely used implementations assume a constant density of states around the Fermi level, which hinders a proper description of materials with distinct features in its vicinity. Here, we present an implementation of the Migdal-Eliashberg theory within the EPW code that considers the full electronic structure and accommodates scattering processes beyond the Fermi surface. To significantly reduce computational costs, we introduce a non-uniform sampling scheme along the imaginary axis. We demonstrate the power of our implementation by applying it to the sodalite-like clathrates YH$_6$ and CaH$_6$, and to the covalently-bonded H$_3$S and D$_3$S. Furthermore, we investigate the effect of maximizing the density of states at the Fermi level in doped H$_3$S and BaSiH$_8$ within the full-bandwidth treatment compared to the constant-density-of-states approximation. Our findings highlight the importance of this advanced treatment in such complex materials. 
\end{abstract}

\date{January 15, 2024}

\maketitle

\section{Introduction}
Discovering and designing new and technologically relevant superconductors is one of the grand challenges of modern science~\cite{lilia2022}. Conventional superconductivity arises from an intricate interplay between the electrons and the vibrational modes of the lattice, which can be condensed into a single parameter known as the electron-phonon (el-ph) coupling strength $\lambda$. This interaction leads to pairing electrons with opposing spins below the critical temperature \tc{}, creating an energy gap at the Fermi surface and resulting in a zero-resistance superconducting condensate. Since the pioneering work of Bardeen, Cooper, and Schrieffer (BCS)~\cite{bardeen1957a}, advancements in computational and theoretical techniques have allowed accurate calculations of $\lambda$ and fully ab-initio predictions of \tc{}~\cite{giustino2017}. The density-functional theory for superconductors~\cite{SCDFT1,SCDFT2,SCDFT3} and the anisotropic Migdal-Eliashberg theory (AME)~\cite{migdal1958a,eliashberg1960} are state-of-the-art examples of such techniques that have contributed significantly in unraveling the properties of the superconducting states of seminal materials like MgB$_2$~\cite{kortus2001,choi2002,choi2003a} and NbS$_2$~\cite{Heil2017} in unprecedented detail, and in predicting entirely novel classes of superconductors from first principles~\cite{mazin2019}. One of the most topical examples is the class of the high-pressure superhydrides~\cite{flores2020,pickard2020,gao2021}, which have revolutionized the search for high-\tc{} superconductivity by demonstrating that detailed calculations of the electronic structure, phonon dispersion, and el-ph coupling can guide experiments in the search for new superconducting materials. Prominent examples would be LaH$_{10}$, theoretically predicted in 2017~\cite{peng2017,liu2017} and experimentally confirmed two years later~\cite{drozdov2019}, \mbox{YH$_6$~\cite{wang2012,kong2021_YH6_exp,snider2021,troyan2021}} and CaH$_6$~\cite{li2015,li2022,ma2022}, or most recently, LaBeH$_8$, the first successfully synthesized ternary superhydride~\cite{zhang2022,PhysRevLett.130.266001}.

The AME formalism is particularly useful in describing the order parameter of weak and strong coupling superconductors. However, computing the el-ph matrix elements and numerically solving the Eliashberg equations requires extremely dense electron and phonon meshes in the Brillouin zone (BZ) to overcome the strong sensitivity to the sampling of the el-ph scattering processes involving states around the Fermi level~\cite{giustino2007}. The AME implementation of the EPW \mbox{code~\cite{EPW,EPW2}}, which was developed by some of the present authors~\cite{margine2013}, enables the interpolation of a small number of el-ph matrix elements to arbitrary electron and phonon wave vectors in the Bloch representation using maximally localized Wannier functions~\cite{WANN1}. This has helped to bridge the gap between experiments and theory and has been widely used in the last few years to determine, among other superconducting properties, the momentum- and band-resolved superconducting order parameter of various anisotropic \mbox{bulk materials~\cite{gao2015,zheng2017,lucrezi2021,lucrezi2022,kafle2022}}, layered \mbox{compounds~\cite{kafle2020,paudyal2020,lian2022}}, and two-dimensional systems~\cite{margine2014,Margine2016-bh,bekaert2017,zheng2020,petrov2021}.

Despite the extraordinary success of the AME implementation in EPW, its main shortage comes from the assumption that the density of states (DOS) is constant for a finite energy window around the Fermi level $\ef$ (of the order of the Debye energies) where the superconducting coupling occurs~\cite{margine2013}. This approximation, widely employed in literature, is valid for a broad range of compounds but will break down for materials with narrow bands or critical points in the vicinity of $\ef$~\cite{PhysRevB.26.1186,Sano2016}, such as van Hove singularities (VHSs) and Lifshitz transitions.

With the present work, we remedy this shortcoming. Our implementation goes beyond the limitations of the previous approach, by explicitly incorporating scattering processes of electrons with energies and momenta beyond the confines of the Fermi surface. This is made possible through the self-consistent determination of the mass renormalization function, energy shift, and order parameter at every temperature while ensuring the system's charge neutrality (see Supplementary Method~1). The corresponding theoretical considerations and equations are detailed in the methods section and in Supplementary Method~1. As this leads to an increased computational workload, we have also implemented a sparse, non-uniform sampling scheme over the imaginary axis, considerably lowering the number of Matsubara frequencies needed compared to the uniform sampling scheme, which, in practice, highly decreases the computational costs (see methods section). 

We apply this implementation to two different classes of topical superhydrides, the sodalite-like clathrates YH$_6$ and CaH$_6$, and the covalently-bonded H$_3$S and D$_3$S.
Results and discussion are provided under \emph{General applications and benchmarking} and material-specific computational details can be found in the methods section.
Moreover, we present compelling evidence that the commonly used approach of computationally optimizing \tc{} by maximizing the DOS at $\ef$ ($N(\ef)$) is often ineffective, as the reported enhancements in \tc{} are, in fact, artifacts resulting from the constant-DOS approximation.
To support this claim, we conduct a detailed study of electron- and hole-doping effects in H$_3$S and BaSiH$_8$ using our full-bandwidth implementation (see \emph{Application to doped hydrides}). The results shed light on the limitations of the maximizing-$N(\ef)$ strategy, emphasizing the need for a more comprehensive and accurate approach in predicting superconducting properties, as provided with our implementation.

\section{Results}
\subsection{General applications and benchmarking}
\label{scexample}
The emergence of superconductivity at record-breaking temperatures has reignited the hope of achieving superconductivity at ambient conditions~\cite{lilia2022}. Indeed, the discoveries of near-room temperature superconductivity in H$_3$S~\cite{duan2014_H3S,drozdov2015} and LaH$_{10}$~\cite{liu2017,drozdov2019,somayazulu2019} at megabar pressures constitute a new landmark for superconductivity and for the prediction of entirely new materials with advanced functionalities fully ab-initio.

Among the numerous already predicted and experimentally confirmed superhydrides, H$_3$S and D$_3$S have received particular attention from the scientific community. Multiple independent experimental groups employing different characterization techniques have confirmed the existence of the cubic $Im\overline{3}m$-H$_3$S structure at pressures around 100-200\,GPa and its superconducting state near room \mbox{temperature~\cite{durajski2016,einaga2016,capitani2017,PhysRevB.95.140101,nakao2019,mozaffari2019,PhysRevB.101.174511,minkov2020}}. Furthermore, after the prediction of stable body-centered cubic structures of hydrogen that form sodalite-like cages containing Ca~\cite{wang2012} and Y~\cite{li2015} atoms above 150\,GPa, YH$_6$ and CaH$_6$ have also been comprehensively studied by many independent theoretical~\cite{peng2017,liu2017,Heil2019,shao2019} and experimental \mbox{groups~\cite{kong2021_YH6_exp,snider2021,troyan2021,li2022,ma2022}}, paving the way for the search for the holy grail of superconductivity~\cite{di2022, di2021, zhang2022, lucrezi2022, ferreira2023}. Many hydrides exhibit interesting features, such as van Hove singularities near $\ef$ and metallic hydrogen states with strong el-ph coupling~\cite{lilia2022}, which make them a unique condensed matter platform to study superconductivity.

Due to their topical relevance and the amount of experimental data available, we have employed the full-bandwidth method to the sodalite-like clathrates YH$_6$ and CaH$_6$, and to the covalently-bonded H$_3$S and D$_3$S. In the following, we will consider two levels of approximation when solving the AME. The first is the Fermi-surface-restricted approximation (FSR)~\cite{EPW}, which, as discussed in detail in the methods section, assumes that the DOS around $\ef$ is constant. The second, and main object of interest in this work, is the full-bandwidth method (FBW), which takes into account the full energy dependence of the DOS and thus allows for the inclusion of el-ph scattering processes away from $\ef$~\cite{EPW2}. 

Furthermore, our implementation of FBW comes in two different flavors: $(i)$ updating the chemical potential, $\cp$, while solving self-consistently the AME equations to maintain the charge neutrality of the system (referred to as FBW+$\mu$ henceforth); and $(ii)$ keeping the chemical potential fixed (referred to simply as FBW) to lighten the computational load. We will also compare our results for \tc{} to the values given by the commonly employed semi-empirical modified McMillan equation~\cite{mcmillan1968a,allen1975a} and the recently proposed machine-learned SISSO model~\cite{xie2021machine}. 

\subsubsection{\texorpdfstring{YH$_6$}{YH6} and \texorpdfstring{CaH$_6$}{CaH6}}
\begin{figure*}[t]
\centering
	\includegraphics[width=\linewidth]{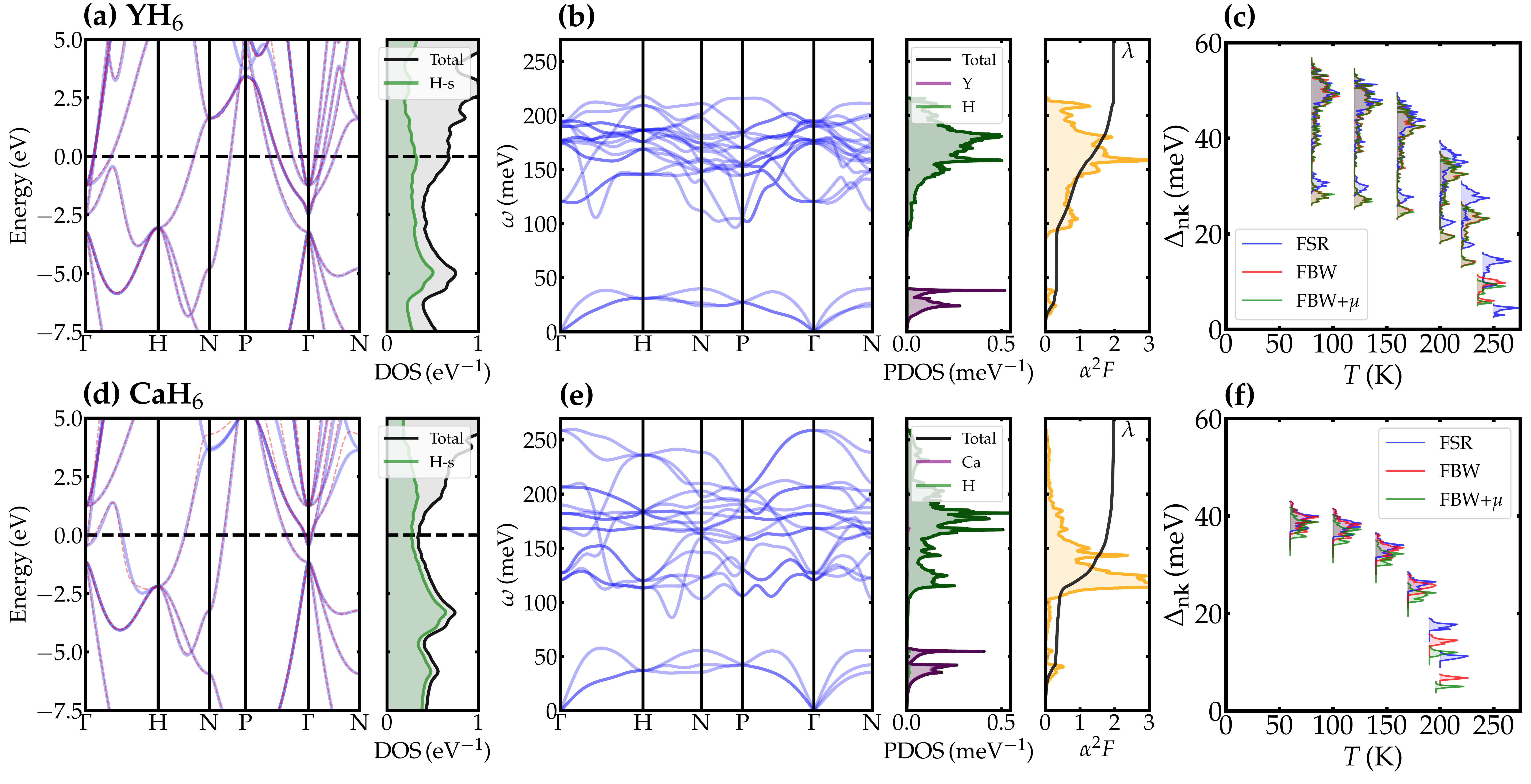}
        \caption{\textbf{Electron, phonon, and superconducting properties for sodalite-like clathrates:} Panel \textbf{(a)} shows the calculated electronic band structure and DOS with respect to the Fermi energy $\ef$ for YH$_6$ at 200\,GPa. The solid blue lines represent the DFT bands, the dashed red lines the Wannier bands, the solid black line the total DOS, the shaded green area the projected DOS for hydrogen s states (H-s), and the dashed black line indicates $\ef$. Panel \textbf{(b)} shows the phonon dispersion (solid blue), the phonon density of states (PDOS, solid black) and its elemental contributions (shaded green and purple), the isotropic Eliashberg spectral function $\alpha^2F(\omega)$ (shaded ochre), and the cumulative electron-phonon coupling parameter $\lambda(\omega)$ (solid black). Panel \textbf{(c)} displays the distribution of the values of the anisotropic superconducting gap \gap\ on the Fermi surface according to the FSR (blue), FBW (red), and FBW+$\mu$ (green) implementations for the Migdal-Eliashberg equations. Panels \textbf{(d)}-\textbf{(f)} show the corresponding results for CaH$_6$ at 200\,GPa.}
 	\label{fig:XH6}
\end{figure*}
To benchmark our implementation, we first take a look at YH$_6$ and CaH$_6$, two hydrides whose variation in the DOS around $\ef$ is rather small, i.e., they are materials for which the FSR approach should be reasonably accurate.

In the following, we summarize the important physical properties to understand the emergence of a high-\tc{} in these materials:
The electronic band structures and DOS for YH$_6$ and CaH$_6$ at a pressure of 200\,GPa are presented in Figs. \ref{fig:XH6}\textbf{(a)} and \textbf{(d)}.
The corresponding phonon dispersions and phonon DOS along with the isotropic Eliashberg spectral function $\alpha^2 F(\omega)$ and the cumulative el-ph coupling strength $\lambda(\omega)$ are reported in Figs.~\ref{fig:XH6}\textbf{(b)} and \textbf{(e)}. 
The relatively high el-ph coupling is associated primarily with the Kohn anomalies observed in the phonon dispersion along the $\Gamma$--$H$ direction for YH$_6 $ and the $H$--$N$ direction for CaH$_6$~\cite{li2015}. For YH$_6$, numerous modes between 90\,meV and 220\,meV significantly contribute to the total el-ph coupling strength. Conversely, the primary contribution for CaH$_6$ is localized in the energy range of 100-160\,meV, derived from the $T_{2g}$ and the $E_g$ modes at the $\Gamma$-zone center belonging to the vibrations of the H$_4$ units~\cite{wang2012}. By integrating $\alpha^2 F(\omega)$, the total el-ph coupling strengths for both YH$_6$ and CaH$_6$ are close to 2.0. This value and the $\alpha^2 F(\omega)$ functions are in excellent agreement with the calculations presented in Refs.~\cite{wang2012,Heil2019}. 

Figures~\ref{fig:XH6}\textbf{(c)} and \textbf{(f)} depict the anisotropic superconducting gap \gap\ as a function of temperature for YH$_6$ and CaH$_6$ at 200\,GPa using the FSR, FBW, and FBW+$\mu$ implementations, with $\cs$ = 0.16. YH$_6$ exhibits two well-defined superconducting gaps on the Fermi surface; a larger, broad energy gap ranging from 35\,meV to 56\,meV at low temperatures, and a smaller gap at approximately 28\,meV, the latter originated from the small zone-centered Fermi surface pockets~\cite{Heil2017}. The gaps of YH$_6$ close at \tc{} $\approx$ 250\,K within the FSR approximation and at \tc{} $\approx$ 238\,K within the FBW approach, independent of whether the chemical potential is updated self-consistently or not. CaH$_6$ is a single-gap superconductor featuring a well-defined gap energy with a maximum of approximately 45\,meV and broadness of about 5\,meV; the \tc{} value is around 200\,K in all three implementations, FSR, FBW, and FBW+$\mu$.

Experimentally, the critical temperature of YH$_6$ varies depending on the synthesis route. Maximum values for \tc{} ranging from 220\,K at 183\,GPa~\cite{kong2021_YH6_exp} to 224\,K at 166\,GPa~\cite{troyan2021} have been reported. In contrast, CaH$_6$ exhibits a wider transition width of approximately 25\,K, with an onset \tc{} of 195\,K at 185\,GPa~\cite{li2022}. A maximum \tc{} of 215\,K for CaH$_6$ has been observed at 172\,GPa~\cite{ma2022}. At 200\,GPa experiments report a \tc{} of about 211\,K for YH$_6$ and about 204\,K for CaH$_6$. Considering the considerable variation of \tc{} with respect to different samples and the uncertainties for a particular measurement, the agreement between experimental values and our numerically determined ones is quite satisfactory. We expect that by including corrections from quantum anharmonic effects, which can be sizeable in hydrides~\cite{errea2016,Errea2020-dz,lucrezi2022_quantum}, the difference between experiment and theory could be even further reduced, which will be a topic of future investigation.

As mentioned before, both YH$_6$ and CaH$_6$ exhibit a slowly varying DOS around $\ef$ and in such cases, the FSR approximation is reasonable for describing the el-ph scattering process around the Fermi level. This is also evident from the fact that the chemical potential remains almost constant throughout the self-consistent solution within FBW+$\mu$. Nevertheless, it is important to note that the FBW implementation offers a better agreement with experiments even for these simple cases.

\subsubsection{\texorpdfstring{H$_3$S}{H3S} and \texorpdfstring{D$_3$S}{D3S}}
\begin{figure*}[ht]
\centering
	\includegraphics[width=\linewidth]{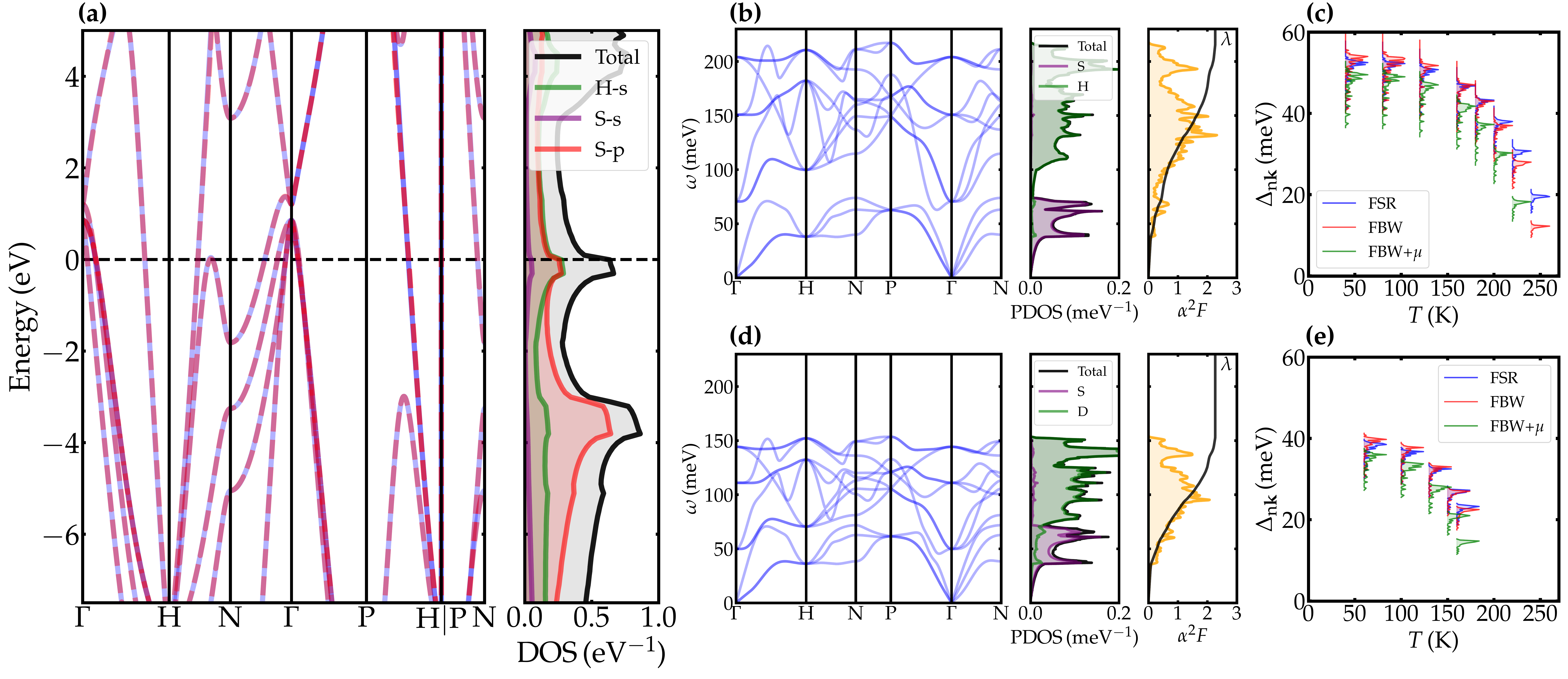}
        \caption{\textbf{Electron, phonon, and superconducting properties for the covalently-bonded materials:} Panel \textbf{(a)} shows the calculated electronic band structure and DOS with respect to the Fermi energy $\ef$ for H$_3$S and D$_3$S at 200\,GPa. The solid blue lines represent the DFT bands, the dashed red lines the Wannier bands, the solid black line the total DOS, the shaded coloured areas the projected DOS for hydrogen s (H-s) and sulfur s and p states (S-s, S-p), and the dashed black line indicates $\ef$. Panel \textbf{(b)} shows the phonon dispersion (solid blue), the phonon density of states (PDOS, solid black) and its elemental contributions (shaded green and purple), the isotropic Eliashberg spectral function $\alpha^2F(\omega)$ (shaded ochre), and the cumulative electron-phonon coupling parameter $\lambda(\omega)$ (solid black). Panel \textbf{(c)} displays the distribution of the values of the anisotropic superconducting gap \gap\ on the Fermi surface according to the FSR (blue), FBW (red), and FBW+$\mu$ (green) implementations for the Migdal-Eliashberg equations. Panels \textbf{(d)} and \textbf{(e)} show the corresponding results for D$_3$S at 200\,GPa.}
 	\label{fig:X3S}
\end{figure*}

In contrast to the $X$H$_6$ materials described above, the behavior of the DOS around $\ef$ is quite different in the covalently bonded H$_3$S and D$_3$S hydrides. As indicated in Fig.~\ref{fig:X3S}\textbf{(a)}, the Fermi level is located right at the shoulder of a marked peak in the DOS, which will give rise to considerable differences between the FSR and FBW approaches. 

Figures~\ref{fig:X3S}\textbf{(b)} and \textbf{(d)} report the phonon dispersion, the phonon DOS, $\alpha^2 F(\omega)$, and the cumulative $\lambda(\omega)$ for H$_3$S and D$_3$S at a pressure of 200\,GPa. The corresponding $\alpha^2 F(\omega)$ functions possess two main peaks in both compounds. For H$_3$S, the dominant one is centered around 120\,meV, and the second one, less intense, around 190\,meV. For D$_3$S, the peaks are shifted to lower frequencies, as expected due to the greater mass of deuterium atoms, with maxima centered around 90\,meV and 130\,meV, respectively. In both hydrides, the whole optical spectra from 30-40\,meV to the Debye frequency contribute to the total electron-phonon coupling, which is found to be $\lambda = 2.3$ for H$_3$S and $\lambda = 2.2$ for D$_3$S, respectively. These values and the corresponding $\alpha^2 F(\omega)$ functions are in excellent agreement with those reported in Refs.~\cite{duan2014_H3S,flores2016,Sanna2018a}.

The solution of the AME equations reveals that H$_3$S and D$_3$S are single-gap superconductors with a broad energy gap distribution. Compared to the FSR treatment, the FBW calculation with fixed $\cp$ lowers the gap energy [see Figs.~\ref{fig:X3S}\textbf{(c)} and \textbf{(e)}]. This effect is even more pronounced when updating $\cp$ self-consistently (FBW+$\mu$), as the chemical potential is shifted to higher energies, moving the Fermi level away from the peak of the van Hove singularity. These results emphasize the critical role of the VHS on the superconducting properties of H$_3$S, as has been pointed out in other works as well~\cite{flores2016,Sano2016,quan2016,jarlborg2016,villa2022}.
 
The highest measured \tc{} in the study of \mbox{Drozdov et al.}~\cite{drozdov2015} is 203\,K at 155\,GPa for H$_3$S and 152\,K at 173\,GPa for D$_3$S, with a variation of \tc{} for different samples of up to 15\,K. Samples with better crystallinity for H$_3$S were later obtained by \mbox{Mozaffari et al.}~\cite{mozaffari2019} with a \tc{} of 201\,K at 155\,GPa and a small transition width of 5.5\,K; and also by \mbox{Nakao et al.}~\cite{nakao2019}, where a sharp drop of the resistance was measured at \tc{} = 200\,K at 150\,GPa. \mbox{Minkov et al.}~\cite{minkov2020} have used the same direct in-situ synthesis from elemental S and excess H$_2$ as in Refs.~\cite{mozaffari2019,nakao2019} to obtain better homogeneous samples for D$_3$S, revealing that D$_3$S reached a maximum \tc{} of 166\,K at 157\,GPa, significantly higher ($\approx$ 10\,\% difference) than previously reported values~\cite{drozdov2015}. 

\begin{table*}[t]
\centering
\caption{\textbf{Summary of the obtained superconducting properties for YH$_6$, CaH$_6$, H$_3$S, and D$_3$S:} The table lists the DOS $N(\ef)$ and the H partial DOS $N_\text{H}(\ef)$ at the Fermi level, the electron-phonon coupling parameter $\lambda$, the logarithmic average of the phonon frequencies $\omega_\text{log}$, and the superconducting critical temperature \tc{} according to the modified McMillan formula (\AD), the University of Florida machine-learning model (\ML), the Fermi-surface-restricted approximation (FSR), and the full-bandwidth implementation without (FBW) and with (FBW+$\mu$) the self-consistent chemical potential update scheme, and all experimental critical temperatures \tc$^{\text{exp}}$ at 200\,GPa available in the literature. The \tc$^{\text{exp}}$ of Ref.~\cite{kong2021_YH6_exp} and Ref.~\cite{ma2022} are used as reference for YH$_6$ and CaH$_6$, respectively; and the \tc$^{\text{exp}}$ of Refs.~\cite{drozdov2015,einaga2016,minkov2020} are used as reference for H$_3$S and D$_3$S.}
\label{tab:tc}
\setlength{\tabcolsep}{6pt}
\begin{tabular}{cccccccccc}
\toprule
 & $N(\ef)\ [N_\text{H}(\ef)]$ & $\lambda$ & $\omega_\text{log}$ & \tc$^{\text{\AD}}$ & \tc$^{\text{\ML}}$ & \tc$^{\text{FSR}}$ & \tc$^{\text{FBW}}$ & \tc$^{\text{FBW+$\mu$}}$ & \tc$^{\text{exp}}$ \\
 & (eV$^{-1}$) &  & (meV) & (K) & (K) & (K) & (K) & (K) & (K) \\ 
\midrule
 YH$_6$  & 0.69 [0.32] & 2.0 & 108 & 154 & 202 & 250 & 239 & 238 & 208-214\\
 CaH$_6$ & 0.31 [0.27] & 2.0 & 104 & 148 & 188 & 205 & 200 & 198 & $\sim 204$ \\
 H$_3$S  & 0.53 [0.27] & 2.3 & 108 & 173 & 232 & 256 & 250 & 232 & 172-184\\
 D$_3$S  & 0.53 [0.27] & 2.2 & \hphantom{1}80 & 127 & 166 & 190 & 182 & 170 & 144-148\\
\bottomrule
\end{tabular}
\end{table*}

As can be appreciated in Tab.~\ref{tab:tc}, FBW+$\mu$ performs best in approaching the experimental critical temperatures \tc$^\text{exp}$ among the different implementations for solving the AME equations. For H$_3$S, our calculations provide \tc{} = 232\,K at 200\,GPa, a percentage difference of 23-30\,\% compared to the experimental values of 172-184\,K at 200\,GPa from Refs.~\cite{drozdov2015,einaga2016,minkov2020}. The rather large differences originate from anharmonicity and the quantum motion of the nuclei, which are known to play a crucial role in H$_3$S~\cite{errea2016}. 
Incorporating these effects is, in principle, possible within EPW, as demonstrated in Refs.~\cite{lucrezi2022_quantum,lucrezi2023} for example, but beyond the scope of the current work. Here, it is important to point out that FBW+$\mu$ provides a much better estimate for \tc{} than FSR.
The performance of FBW+$\mu$ in reproducing the experimental values is notably better for D$_3$S, where anharmonic and quantum ionic effects are smaller. The full-bandwidth treatment only slightly modifies the structure of the superconducting gap \gap, i.e., the gap distributions are shifted to lower energies while retaining the overall shape at each temperature. 

The isotope effect coefficient, according to the BCS theory, is given by
\begin{align}
    \alpha = -\dfrac{\ln T_{\text{c}}^{\text{D$_3$S}} - \ln T_{\text{c}}^{\text{H$_3$S}}}{\ln M_{\text{D}} - \ln M_{\text{H}}},
\end{align}
where $M_\text{H}$ and $M_\text{D}$ are the atomic mass of hydrogen and deuterium. The experimental values obtained for $\alpha$ are around 0.47 at 150\,GPa~\cite{durajski2016}. Within FSR, we obtain $\alpha$ = 0.54 at 200\,GPa; for FBW we have $\alpha$ = 0.48; and for FBW+$\mu$ we have $\alpha$ = 0.45. 

These results demonstrate that the full-bandwidth method is imperative for accurately describing these systems within the Eliashberg formalism. As already pointed out by other authors~\cite{flores2016,Sano2016,quan2016}, the conventional Eliashberg formalism within the FSR framework partially fails to accurately describe the \tc{} behavior in H$_3$S due to a substantial variation of the DOS near $\ef$ originating from the van Hove-type singularity present there. The strong el-ph coupling of \mbox{$\lambda$ = 2.3} for H$_3$S and \mbox{$\lambda$ = 2.2} for D$_3$S, and the broad distribution of $\alpha^2 F$ over the vibrational spectra makes this scenario even more dramatic since the region around $\ef$, where the phonon scattering dominates over the Coulomb repulsion, is strongly enhanced by the el-ph interactions. Lastly, these results highlight the importance of updating the chemical potential while solving the AME equations within the FBW treatment. 

At the end of this section, we also want to shortly discuss the modified McMillan~\cite{mcmillan1968a,allen1975a} formula and a recent machine-learning approach to improve upon it~\cite{xie2021machine}, both of which can be used as an almost instant way to predict \tc{}, but do not offer insight into other properties of the superconducting state. The modified McMillan formula (\AD, see Supplementary Note 1) is obtained from the Eliashberg theory by defining moments of the $\alpha^2 F$ spectral function and fitting an equation to match the experimental \tc{}, taking into account a bit more than 200 data points. As the data set consisted of the (low-\tc{}) superconductors known at the time and a few pure model calculations, it does not reproduce the experimental critical temperatures of the extreme cases of contemporary highly-compressed high-\tc{} materials, as is also evident for the materials chosen in this study~\cite{allen1975a}.

The machine-learned equation proposed by a group of the University of Florida (\ML), on the other hand, has been trained to match the solutions of the Migdal-Eliashberg equations with a dataset of thousands of real and artificially generated $\alpha^2 F$ functions, including high-pressure, high-\tc{} hydrides as well. It thus performs much better than the \AD{} formula and matches the results obtained with AME fairly well. It is therefore a viable tool to determine an accurate value for \tc{} quickly but does not offer insight into the superconducting gap function and its energy distribution, the superconducting DOS, and so on. We also observe shortcomings when trying to simulate the effects of doping, as detailed in the next section.

\subsection{Application to doped hydrides}
\label{sec:doping}
Doping can be a powerful method to tailor and fine-tune specific material properties, especially in electronic applications.
It has been used to metallize semiconducting phases, induce superconductivity, and optimize specific properties of the superconducting phase~\cite{kim2007,yazici2013,margine2014,lian2022,bhattacharyya2022,correa2023}.
In particular, this approach has been employed to increase the \tc\ in known superconducting systems and is claimed to be a route to obtain (or to be the source of) room-temperature superconductivity in recently reported hydrides~\cite{sun2019,olea2019,guan2021,wang2022,ge2021,cataldo2023,dasenbrock2023,ferreira2023}.

The doping of hydrogen-rich superconductors and its tentatively beneficial effect on \tc\ has been extensively studied in various works~\cite{cui2020,sun2020,ge2020,guan2021,doi:10.7566/JPSJ.87.124711,PhysRevB.93.224513,shao2022,Heil2015,FAN2016105,VILLACORTES20222333,Villa_Cortes_2022,wang2022,feng2022}, however, most of the theoretical predictions for an enhancement of \tc\ rely on calculations based on the McMillan/Allen-Dynes formulas, or on the ME equations within the constant-DOS approximation, and hence focus mainly on maximizing the value of the DOS at the Fermi level and thereby that of $\lambda$. In particular, in systems with VHS-like peaks close to the Fermi level, this approach might severely overestimate the contribution of the electronic states available for the superconducting pairing and thus also the critical temperature. Furthermore, the exact value of the DOS close to a VHS-like shape is subject to large variations, which makes predictions even more error-prone~\cite{quan2016}.

\begin{figure*}
\centering
\includegraphics[width=\textwidth]{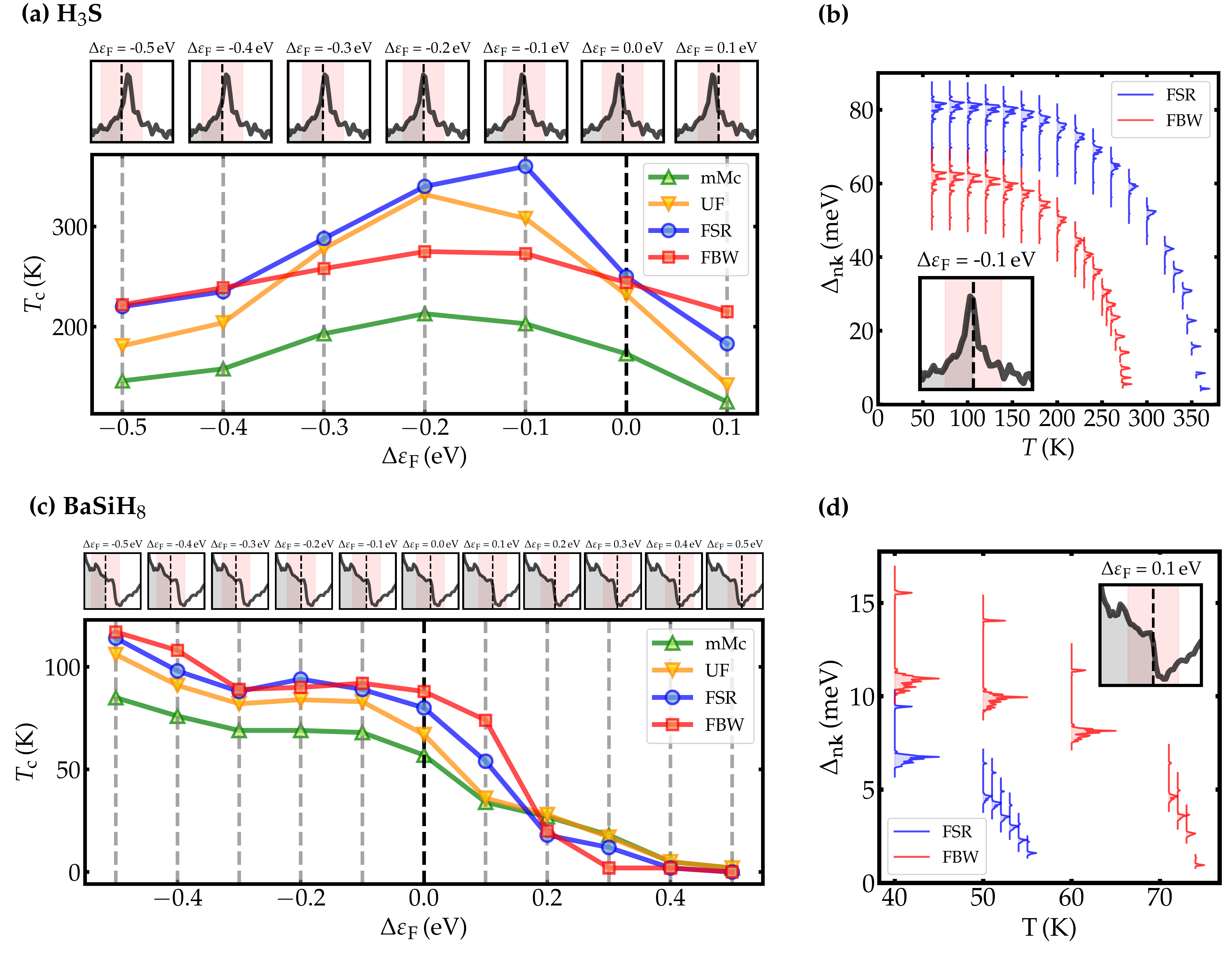}
\caption{\textbf{Doping effects on the critical temperature in different approaches:} Effects of doping (shifting the Fermi level) in H$_3$S at 200\,GPa (panels \textbf{(a)} and \textbf{(b)}) and BaSiH$_8$ at 30\,GPa (panels \textbf{(c)} and \textbf{(d)}). Panel \textbf{(a)} shows the superconducting critical temperature \tc{} as a function of the Fermi level shift $\Delta\ef$, obtained within the \AD{} formula (green), the \ML{} equation (orange), the FSR approximation (blue), and the FBW approach (red). The smaller subpanels on top show the corresponding DOS in a range of $\pm\SI{2}{eV}$ around the unshifted Fermi energy $\ef$, where the dashed lines mark the position of $\ef+\Delta\ef$ and the shaded red areas highlight the included electronic energy range of $\ef+\Delta\ef\pm\SI{1}{eV}$. Panel \textbf{(b)} displays the distribution of the values of the anisotropic superconducting gap \gap\ on the Fermi surface within the FSR (blue) and FBW (red) approach for H$_3$S with $\Delta\ef$ = -0.1\,eV, where we find the maximum absolute difference between the FSR and FBW \tc{} (see blue and red lines in \textbf{(a)}). The inset shows the corresponding DOS subpanel from \textbf{(a)}. Panels \textbf{(c)} and \textbf{(d)} show the corresponding results for BaSiH$_8$ at 30\,GPa, where we find the maximum \tc{} difference for  $\Delta\ef$ = +0.1\,eV.}
\label{fig:doping}
\end{figure*}

With the FBW implementation, we overcome these problems and limitations as demonstrated for H$_3$S and BaSiH$_8$, two materials exhibiting considerable variation in the DOS around the Fermi level. Applying doping via a rigid band shift, we explore the different regions of increased, maximal, and lowered DOS, and discuss the effects on \tc\ within the \AD{} and MD formulas, and the FSR and FBW methods. We want to note here that for these calculations we do not update the chemical potential, as an explicit shift in the Fermi level can also be interpreted as a shift in the chemical potential, and hence such a calculation can be reproduced by one with a (slightly) different effective shift.

\subsubsection{\texorpdfstring{H$_3$S}{H3S}}
The well-studied hydride superconductor H$_3$S is a perfect candidate material where doping to optimize \tc{} appears very tempting due to the VHS-like peak in the DOS in close proximity to the Fermi level, as shown in Fig.~\ref{fig:doping}\textbf{(a)}-\textbf{(b)}.
Many theoretical works have considered doping of H$_3$S to increase its \tc\ estimates, choosing dopants to bring the system's Fermi level closer to the maximum of the VHS-like peak. The stability of the $Im\overline{3}m$ H$_3$S structure with incorporation of various elements has been systematically investigated using either the direct supercell approach with substitutional doping or the virtual crystal approximation, followed by \tc\ estimates based on the \AD{} formula or the isotropic FSR \mbox{approach~\cite{PhysRevB.93.224513,ge2020,PhysRevB.98.174101,li2018,shao2022,FAN2016105,feng2022,cui2020,sun2020,guan2021,doi:10.7566/JPSJ.87.124711}}. 

While adding dopants may increase the \tc\ by also enhancing the el-ph coupling, we want to specifically address the effect of a change in the number of available electronic states. To this end, we solved the AME equations in the FSR and FBW frameworks for shifts of the Fermi level $\Delta\ef$ between \SI{-0.5}{eV} and \SI{+0.1}{eV} in steps of \SI{0.1}{eV}, corresponding to changes in the electron number $\Delta_\text{elec}$ of -0.30, -0.26, -0.20, -0.13, -0.06, and +0.05. 

In Fig.~\ref{fig:doping}\textbf{(a)}, we present the \tc\ values obtained within the FSR and FBW treatments of the AME equations and the results obtained using the \AD{} and \ML{} semi-empirical formulas mentioned earlier. The \tc\ values are plotted as a function of $\Delta\ef$ demonstrating that all approaches show a clear correlation between \tc\ and DOS with the maximum \tc\ occurring at around $\Delta\ef \sim$ \SI{-0.1}{eV} to \SI{-0.2}{eV}, as highlighted in Fig.~\ref{fig:doping}\textbf{(b)}.

However, within the FSR approximation, there is a notable and unphysical increase in \tc{} values around the maximum of the DOS. This behavior can be attributed to the limitation of the constant-DOS assumption in the FSR approach. Similarly, the \ML{} model exhibits a strong dependence on doping.
In contrast, the AME solutions in the FBW approach, which consider the full energy dependence of the DOS, exhibit a much less pronounced effect of doping on \tc{}.
Finally, the semi-empirical \AD{} formula consistently underestimates \tc{} for strongly coupled systems when compared to results obtained from the AME equations. 

\begin{figure}
	\includegraphics[width=\columnwidth]{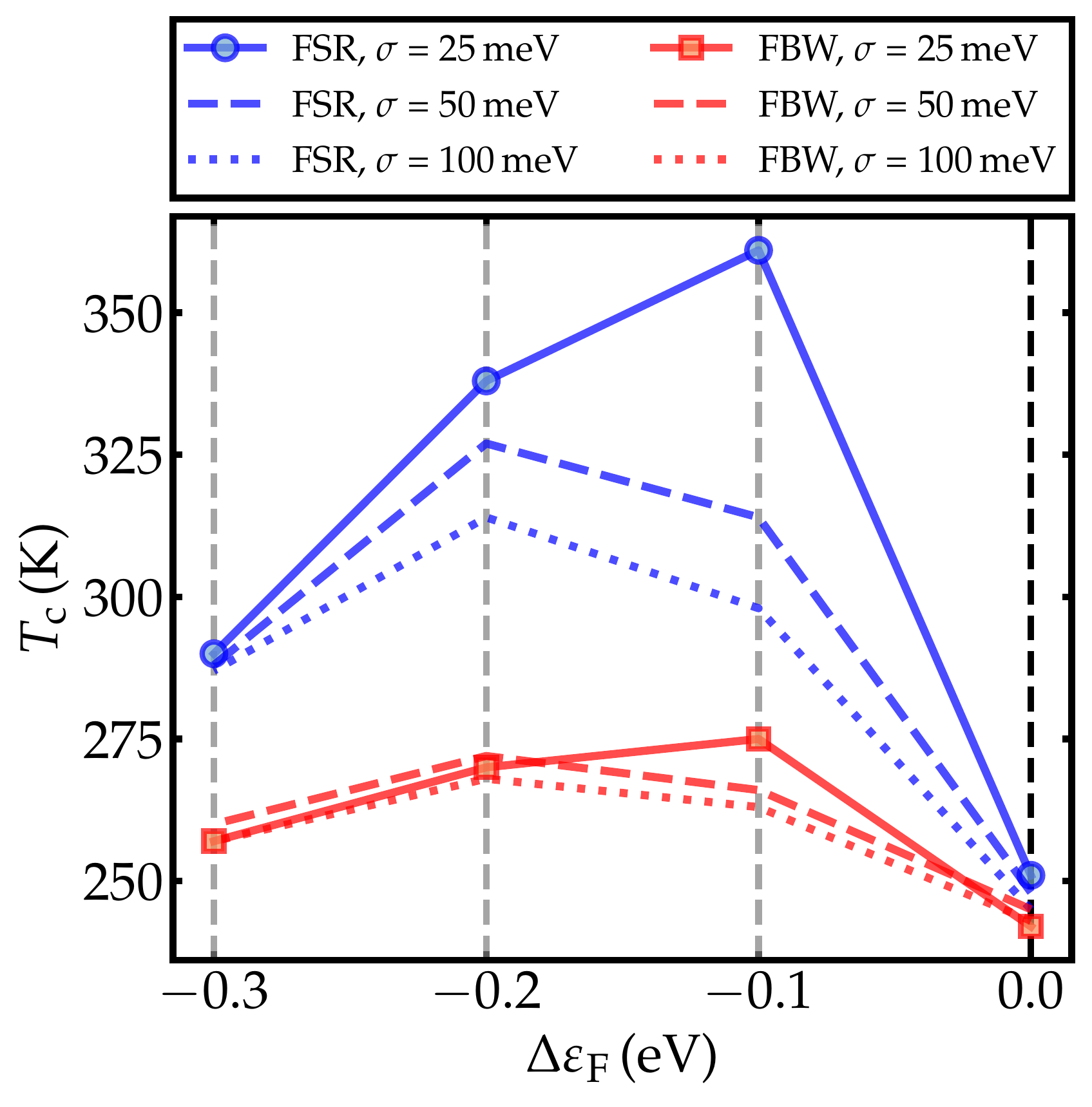}
	\caption{\textbf{Dependence of the doped \tc{} results on the electronic smearing:} Influence of the smearing value $\sigma$ for the energy-conserving $\delta$-functions on the \tc\ of doped H$_3$S within the FSR (blue) and FBW (red) approach as a function of the Fermi energy shift $\Delta\ef$. The solid lines represent the corresponding results shown in Figure~\ref{fig:doping}\textbf{(a)} for a smearing value of $\sigma=\SI{25}{meV}$, the dashed (dotted) lines represent the results for $\sigma=50\ (100)\,\text{meV}$.}
	\label{fig:tc_fsr_fbw_sm_H3S}
\end{figure}

In the context of our analysis of a doped H$_3$S system, we would like to address the topic of speculated room temperature superconductivity within this system~\cite{snider2020room}. Considering that the experimental \tc{} values for pure H$_3$S are around or below 200\,K, and that the FBW calculations only show a maximum increase in \tc{} of about 5-10\% upon doping (see Fig.~\ref{fig:tc_fsr_fbw_sm_H3S}), achieving a conventional superconducting state at room temperature in the H$_3$S parent phase remains elusive; at least within the assumption that the slight doping leaves the electronic structure unaltered.

On a more technical note, we want to add that VHS-like features also pose other problems when trying to arrive at robust numerical results. For example, the exact value of the DOS at the VHS-like peak is difficult to converge and requires extremely dense \textbf{k}-grids and small smearing values. To demonstrate this, we examined the VHS region by performing additional calculations with two different smearing values $\sigma$ for the energy-conserving $\delta$-functions. The results are shown in Figure~\ref{fig:tc_fsr_fbw_sm_H3S}.
The critical temperature obtained with the FSR approximation is very sensitive to the smearing value chosen, the largest deviation in \tc{} reaching up to 70\,K, whereas the FBW implementation is considerably more robust, with at most a 10\,K difference between the different smearing values tested. The reason for that is that the only quantity impacted by smearing in the FBW implementation is the screened Coulomb interaction term in Eq.~(\ref{xsceq18}), whereas smearing impacts all FSR equations via $N(\ef)$ and $\delta(\varepsilon_{m \mathbf{k}+\mathbf{q}} - \ef)$ (see Eqs.~\ref{xsceq21} and \ref{xsceq22}). As a result, besides being more rigorous, the FBW implementation also considerably improves convergence behavior for materials exhibiting a strongly varying DOS.

\subsubsection{\texorpdfstring{BaSiH$_8$}{BaSiH8}}
The $Fm\overline{3}m$ phase of BaSiH$_8$ was recently predicted and described in detail by some of the current authors~\cite{lucrezi2022,lucrezi2022_quantum}.
This ternary hydride has the crystal structure of the $XY$H$_8$ template first introduced for LaBH$_8$~\cite{di2021}, also assumed by other high-\tc\ superhydrides that are stable at moderate \mbox{pressures~\cite{zhang2022,hou2022}}, like the recently synthesized LaBe-compound~\cite{PhysRevLett.130.266001}. Up to now, BaSiH$_8$ is the $XY$H$_8$ compound with the lowest predicted pressure of dynamical stability of \SI{5}{GPa} within the harmonic phonon theory. A more realistic estimate for the critical pressure of stability and synthesizability is provided by considering the kinetic stability of the compound, which places stability at a pressure above \SI{30}{GPa}~\cite{lucrezi2022,lucrezi2022_quantum}.

Independent of the actual critical pressure, this material exhibits a step-like feature around the Fermi level, with an almost constant region of high DOS below the Fermi level and a sharp drop to a region of very low DOS above $\ef$, as can be appreciated in Fig.~\ref{fig:doping}\textbf{(c)}, making it another perfect test bed to compare the FSR and FBW approaches.

We solved the AME equations within FSR and FBW approaches for shifts of the Fermi level $\Delta\ef$ between \SI{-0.5}{eV} and \SI{+0.5}{eV} in steps of \SI{0.1}{eV}, corresponding to changes in the electron number $\Delta_\text{elec}$ of -0.52, -0.42, -0.31, -0.21, -0.10, +0.09, +0.15, +0.18, +0.20, and +0.22.
In Figs.~\ref{fig:doping}\textbf{(c)} and \textbf{(d)}, we show the \tc\ values obtained for the two levels of approximations as a function of $\Delta\ef$. As before, \tc\ roughly follows the shape of the DOS, but, in contrast to H$_3$S, we observe a considerable increase in \tc{} for BaSiH$_8$ when employing the more rigorous FBW approach. Already in the undoped case the \tc{} is raised to 87\,K, which can be further increased by doping to about 92\,K when shifting the Fermi level by \mbox{-0.1\,eV}. 
This would place the critical superconducting temperature of BaSiH$_8$ above the technologically extremely important threshold set by the boiling temperature of nitrogen. 

The behaviour of \tc{} with respect to Fermi level shifts for the different methods is quite complex: For a shift of -0.1\,eV or below the AME results are similar and also agree with the \ML{} model, while \AD{} gives considerably lower values for \tc. In a region around 0.1\,eV shift, the FBW implementation predicts a larger value for \tc{} than FSR, and the simple \AD{} and \ML{} formulas provide an even smaller estimate. For $\ef$-shifts larger than 0.2\,eV, \AD{} and \ML{} actually give the largest values for \tc{} and FBW the smallest. In other words, while the dependence of \tc{} with respect to Fermi level shifts was smallest in FBW for H$_3$S, it is actually varying the strongest within FBW for BaSiH$_8$.

These intricate results underscore the importance of taking into account the energy dependence of the electronic DOS around the Fermi level, in particular for systems where the DOS is either strongly peaked close to \mbox{$\ef$~\cite{wang2021,wang2022,durajski2023,cataldo2023,jiang2023}}, as for H$_3$S, or highly asymmetric~\cite{song2021,zhong2022,di2023,ma2023,lucrezi2023}, as for BaSiH$_8$. In that light, we believe that the agreement between experimental measurements and theoretical predictions can be considerably improved by employing the FBW method not only for the class of superhydrides but also for other material systems showing similar features in the \mbox {DOS~\cite{margine2014,ferreira2018,zheng2020,liu2022,gai2022,ding2022,bekaert2022,sevik2023,hao2023,liu2023,luo2023}}.

\section{Discussion}
In summary, we have employed the anisotropic Migdal-Eliashberg formalism within the full-bandwidth formulation utilizing maximally-localized Wannier functions as implemented in the EPW suite. This approach enables us to calculate the momentum- and band-resolved superconducting gap more accurately, taking into account the electron-phonon scattering processes around the Fermi level, and not only restricted to the Fermi surface. In addition, we introduced a sparse, non-uniform sampling scheme over the imaginary Matsubara frequencies, which shows similar accuracy and much-improved efficiency compared to the uniform sampling scheme.

To validate the robustness of our methodology, we conducted comprehensive tests on two representative classes of superhydrides: the sodalite-like clathrates YH$_6$ and CaH$_6$, as well as the covalent hydrides H$_3$S and D$_3$S. To assess the accuracy of our approach, we compared our results with previous ab-initio calculations and experimental data.
Our results unequivocally demonstrate the indispensable role of the full-bandwidth formulation, particularly for compounds characterized by narrow bands or critical points in proximity to the Fermi level. A noteworthy illustration of the importance of employing the FBW formulation is evident in the case of H$_3$S, which possesses a van Hove singularity. Our methodology effectively captures the intricate behavior of such systems, highlighting the superiority of the FBW approach in these critical scenarios.
Furthermore, we emphasize the crucial impact of the chemical potential updating scheme within the FBW formulation, which substantially contributes to accurately describing the superconducting phase in these challenging cases. This aspect proves to be a vital component in achieving a comprehensive understanding of the superconducting properties of these complex materials. 

In addition, we applied the FBW approach to investigate electron- and hole-doped hydride superconductors, namely H$_3$S and the recently predicted low-pressure BaSiH$_8$. These materials serve as prime examples of systems with distinct DOS features that deviate significantly from the constant-DOS assumption made in the FSR approximation.
Previous studies have often focused on maximizing the DOS specifically at the Fermi level, aiming to design high(er)-\tc\ superconductors by doping the system to shift $\ef$ to the maximum of a VHS-like structure. Our calculations reveal significant pitfalls associated with such a simplistic FSR-based approach. We find instances of pronounced under- or overestimation of \tc, highlighting the critical importance of adopting the FBW method, particularly in scenarios with strongly peaked DOS or closely adjacent DOS regions exhibiting extremely high and low values.
By employing the FBW approach, we can accurately capture the intricate interplay of DOS features and better predict the behavior of superconductors in these complex cases. This sheds light on the limitations of the FSR approach and underscores the significance of our advanced methodology in studying and engineering novel superconducting materials with tailored properties for real-world applications.

\section{Methods}
\label{sec:theory}

\subsection{Anisotropic Migdal-Eliashberg theory}
\label{sec:theory_aniso}
The Eliashberg theory is a powerful many-body perturbation approach for describing conventional superconductors, where the Cooper pairing between two electrons stems from the interplay between the attractive el-ph coupling and the repulsive screened Coulomb interaction~\cite{eliashberg1960,eliashberg1961}. 
The Nambu-Gor'kov's formalism~\cite{gorkov1958a,nambu1960a} with a generalized matrix Green's function can be used to formulate the Eliashberg theory. The on- and off-diagonal elements of the $2\times2$ Green's function matrix describe the single-particle excitations in the normal state and Cooper-pair amplitudes in the superconducting state, corresponding to the standard and anomalous Green’s functions, respectively~\cite{margine2013,gorkov1958a,nambu1960a,garland1967a,allen1983b,carbotte1990a,choi2003a}. The transition from normal to superconducting state manifests in anomalous Green's functions becoming nonzero below a material specific critical temperature.

The matrix Green's function is obtained from the Dyson equation 
\begin{eqnarray}\label{xsceq4}
\hat{G}_{n \mathbf{k}}^{-1}(i \omega_{j})=
\left[\hat{G}_{n \mathbf{k}}^{0}(i \omega_{j})\right]^{-1}
-\hat{\Sigma}_{n \mathbf{k}}^{\mathrm{pa}}(i \omega_{j}),
\end{eqnarray}
where $\hat{G}_{n \mathbf{k}}^{0}(i \omega_{j})$ is the non-interacting Green’s function in the normal state with band index $n$ and wavevector $\mathbf{k}$, $\hat{\Sigma}_{n \mathbf{k}}^{\mathrm{pa}}(i \omega_{j})$ is the pairing self-energy, and $i\omega_j = i(2j + 1)\pi T$ is the fermionic Matsubara frequency with $T$ being the absolute temperature and $j$ an integer. An expression for the self-energy in terms of the electron Green's function can be obtained with the el-ph and electron-electron contributions given by the \mbox{Migdal~\cite{migdal1958a,allen1983b}} and GW~\cite{hedin1965s,hybertsen1986s} approximations, respectively:
\begin{eqnarray}\label{xsceq5}
\hat{\Sigma}_{n \mathbf{k}}^{\mathrm{pa}}(i \omega_{j})=
\hat{\Sigma}_{n \mathbf{k}}^{\mathrm{ep}}(i \omega_{j})+
\hat{\Sigma}_{n \mathbf{k}}^{\mathrm{c}}(i \omega_{j}).
\end{eqnarray}
Using the Pauli matrices,
$\hat{\tau}_{0}=\bigl( \begin{smallmatrix}1 &  0\\ 0 &  1\end{smallmatrix}\bigr),
 \hat{\tau}_{1}=\bigl( \begin{smallmatrix}0 &  1\\ 1 &  0\end{smallmatrix}\bigr),
 \hat{\tau}_{2}=\bigl( \begin{smallmatrix}0 & -i\\ i &  0\end{smallmatrix}\bigr),
 \hat{\tau}_{3}=\bigl( \begin{smallmatrix}1 &  0\\ 0 & -1\end{smallmatrix}\bigr)$,
the el-ph and Coulomb contributions to the self-energy can be expressed as
\begin{eqnarray}\label{xsceq6}
\hat{\Sigma}_{n \mathbf{k}}^{\mathrm{ep}}(i \omega_{j}) =
-T \sum_{m j^{\prime} \nu} \int\!\frac{\mathrm{d} \mathbf{q}}{\Omega_{\rm BZ}}\,
\hat{\tau}_{3} 
\hat{G}_{m \mathbf{k}+\mathbf{q}}(i \omega_{j^{\prime}}) \hat{\tau}_{3} \nonumber\\
 \times \, \left|g_{m n \nu}(\mathbf{k}, \mathbf{q})\right|^{2} 
D_{\mathbf{q} \nu}(i \omega_{j}-i \omega_{j^{\prime}}),
\end{eqnarray}
and
\begin{eqnarray}\label{xsceq7}
\hat{\Sigma}_{n \mathbf{k}}^{c}(i \omega_{j})=
-T \sum_{m j^{\prime}} \int\!\frac{\mathrm{d} \mathbf{q}}{\Omega_{\rm BZ}}\, \hat{\tau}_{3} 
\hat{G}^{\rm od}_{m \mathbf{k}+\mathbf{q}}(i \omega_{j^{\prime}}) 
\hat{\tau}_{3} V_{n \mathbf{k},m \mathbf{k}+\mathbf{q}},
\end{eqnarray}
where $\Omega_{\rm BZ}$ is the BZ volume, $D_{\mathbf{q}\nu}(i\omega_{l})=2\omega_{\mathbf{q}\nu}/[(i\omega_{l})^2-\omega_{\mathbf{q} \nu}^2]$ is the dressed phonon propagator for phonons with wavevector $\mathbf{q}$ and branch index $\nu$, $i\omega_{l}=i2l\pi T$ is the bosonic Matsubara frequency with $l$ an integer, $g_{m n \nu}(\mathbf{k}, \mathbf{q})$ is the screened el-ph matrix element for the scattering between the electronic states $n \mathbf{k}$ and $m \mathbf{k}+\mathbf{q}$ through a phonon of frequency $\omega_{\mathbf{q} \nu}$, and 
$V_{n \mathbf{k},m \mathbf{k}+\mathbf{q}}$ is the static screened Coulomb interaction between electrons~\cite{allen1983b,marsiglio2008a}. In Eq.~(\ref{xsceq7}), only the off-diagonal
components of the Green’s function $\hat{G}^{\rm od}_{n \mathbf{k}}(i \omega_{j})$ are retained in order to avoid double counting the Coulomb effects that are already included in $\hat{G}_{n \mathbf{k}}^{0}(i \omega_{j})$~\cite{allen1983b}.

The anisotropic el-ph coupling strength is described as
\begin{eqnarray}\label{xsceq8}
\lambda_{n \mathbf{k}, m \mathbf{k}+\mathbf{q}} (\omega_{j}-\omega_{j^{\prime}}) =
N(\ef)
\sum_{\nu}
\frac{2 \omega_{\mathbf{q} \nu} \left|g_{m n \nu}(\mathbf{k}, \mathbf{q})\right|^{2} }{(\omega_{j}-\omega_{j^{\prime}})^{2}
+\omega_{\mathbf{q} \nu}^{2}},
\end{eqnarray}
with $N(\ef)$ the DOS per spin at the Fermi level. Eq.~(\ref{xsceq8}) can be used to rewrite the el-ph self-energy in Eq.~(\ref{xsceq6}), which can then be taken together with Eq.~(\ref{xsceq7}) and inserted into Eq.~(\ref{xsceq5}). The pairing self-energy then becomes:
\begin{eqnarray}\label{xsceq10}
\hat{\Sigma}_{n \mathbf{k}}^{\mathrm{pa}}(i \omega_{j})&=& 
\frac{T}{N(\ef)} \sum_{m j^{\prime}} \int\!\frac{\mathrm{d} \mathbf{q}}{\Omega_{\rm BZ}} \nonumber \\
&& \hspace{-1.5cm} \times \, \Bigg\{ 
\lambda_{n \mathbf{k}, m \mathbf{k}+\mathbf{q}} (\omega_{j}-\omega_{j^{\prime}}) 
 \hat{\tau}_{3} \hat{G}_{m \mathbf{k}+\mathbf{q}}(i \omega_{j^{\prime}}) \hat{\tau}_{3} \nonumber \\
&& \hspace{-1.1cm} - \, N(\ef) V_{n \mathbf{k},m \mathbf{k}+\mathbf{q}} 
\hat{\tau}_{3} \hat{G}^{\rm od}_{m \mathbf{k}+\mathbf{q}}(i \omega_{j^{\prime}}) \hat{\tau}_{3} \Bigg\}
\end{eqnarray}
To replace $\hat{G}_{n \mathbf{k}}(i \omega_{j})$ in Eq.~(\ref{xsceq10}), we expand the two components of the Dyson equation (\ref{xsceq4}) in terms of the Pauli matrices as
\begin{equation}\label{xsceq11}
\left[\hat{G}_{n \mathbf{k}}^{0}(i \omega_{j})\right]^{-1} = 
i \omega_{j} \hat{\tau}_{0}-(\varepsilon_{n \mathbf{k}}- \cp) \hat{\tau}_{3},
\end{equation}
where $\varepsilon_{n \mathbf{k}}$ are the Kohn-Sham eigenenergies, and
\begin{eqnarray}\label{xsceq12}
\hat{\Sigma}_{n \mathbf{k}}^{\mathrm{pa}} (i \omega_{j}) &=&
i \omega_{j}\left[1-Z_{n \mathbf{k}}(i \omega_{j})\right] \hat{\tau}_{0}+\chi_{n \mathbf{k}}(i \omega_{j}) \hat{\tau}_{3}\nonumber\\ 
&+&\phi_{n \mathbf{k}}(i \omega_{j}) \hat{\tau}_{1}+
\bar{\phi}_{n \mathbf{k}}(i \omega_{j}) \hat{\tau}_{2}.
\end{eqnarray}
In Eq.~(\ref{xsceq12}), we introduced the mass renormalization function $Z_{n \mathbf{k}}(i \omega_{j})$, the energy shift $\chi_{n \mathbf{k}}(i \omega_{j})$, and the order parameter $\phi_{n \mathbf{k}}(i \omega_{j})$. Inserting Eqs.~(\ref{xsceq11}) and (\ref{xsceq12}) into Eq.~(\ref{xsceq4}), and inverting the resulting matrix leads to the following expression for $\hat{G}_{n \mathbf{k}}(i \omega_{j})$:
\begin{eqnarray}\label{xsceq13}
\hat{G}_{n \mathbf{k}} (i \omega_{j}) = \frac{1}{\det[\hat{G}_{n \mathbf{k}}^{-1}(i \omega_{j})]}
\Bigg\{ i \omega_{j} Z_{n \mathbf{k}}(i \omega_{j}) 
\hat{\tau}_{0} \nonumber\\
+ \Bigl[ \varepsilon_{n \mathbf{k}}-\cp+
\chi_{n \mathbf{k}}(i \omega_{j}) \Bigr] \hat{\tau}_{3}
+\phi_{n \mathbf{k}}(i \omega_{j}) \hat{\tau}_{1}+
\bar{\phi}_{n \mathbf{k}}(i \omega_{j}) \hat{\tau}_{2}\Bigg\} \nonumber \\
\end{eqnarray}
It can be easily verified that $\phi_{n \mathbf{k}}(i \omega_{j})$ and $\bar{\phi}_{n \mathbf{k}}(i \omega_{j})$ are proportional within an arbitrary phase, and without loss of generality, one can choose the relative phase such that \mbox{$\bar{\phi}_{n \mathbf{k}}(i \omega_{j})=0$~\cite{allen1983b,EPW}}.
Eq.~(\ref{xsceq13}) with $\bar{\phi}_{n \mathbf{k}}(i \omega_{j})=0$ can be used to rewrite Eq.~(\ref{xsceq10}) for the pairing self-energy as:
\begin{eqnarray}\label{xsceq14}
\hat{\Sigma}_{n \mathbf{k}}^{\mathrm{pa}} (i \omega_{j})=
-\frac{T}{N(\ef)} \sum_{m j^{\prime}} \int\!\frac{\mathrm{d} \mathbf{q}}{\Omega_{\rm BZ}}
\frac{1}{\theta_{m \mathbf{k}+\mathbf{q}}(i \omega_{j^{\prime}})} 
\nonumber \\
\times \, \Bigg\{i \omega_{j^{\prime}} Z_{m \mathbf{k}+\mathbf{q}}(i \omega_{j^{\prime}}) 
 \lambda_{n \mathbf{k}, m \mathbf{k}+\mathbf{q}}( \omega_{j}-\omega_{j^{\prime}}) \hat{\tau}_{0} 
\nonumber \\
 + \, \Bigl[ \varepsilon_{m \mathbf{k}+\mathbf{q}}-\cp +\chi_{m \mathbf{k}+\mathbf{q}}(i \omega_{j^{\prime}})\Bigr]
 \lambda_{n \mathbf{k}, m \mathbf{k}+\mathbf{q}}( \omega_{j}-\omega_{j^{\prime}}) \hat{\tau}_{3} 
\nonumber \\
 - \, \phi_{m \mathbf{k}+\mathbf{q}}(i \omega_{j^{\prime}})\left[ \lambda_{n \mathbf{k}, m \mathbf{k}+\mathbf{q}}( \omega_{j}-\omega_{j^{\prime}}) 
 - N(\ef)V_{n \mathbf{k},m \mathbf{k}+\mathbf{q}} \right] 
 \hat{\tau}_{1} \Bigg\}, 
\nonumber\\
\end{eqnarray}
where
\begin{eqnarray}\label{xsceq15}
\theta_{n \mathbf{k}}(i \omega_{j})
&=&-\det[\hat{G}_{n \mathbf{k}}^{-1}(i \omega_{j})]|_{\bar{\phi}_{n \mathbf{k}}(i \omega_{j})=0} \nonumber\\
&=&
\left[\omega_{j} Z_{n \mathbf{k}} (i \omega_{j})\right]^{2}
+\left[\varepsilon_{n \mathbf{k}}- \cp+\chi_{n \mathbf{k}}(i \omega_{j})\right]^{2}\nonumber \\
&+&\left[\phi_{n \mathbf{k}}(i \omega_{j})\right]^{2}. 
\end{eqnarray}
Equating the different components of the Pauli matrix elements in Eqs.~\eqref{xsceq12} and \eqref{xsceq14} leads to a system of three coupled non-linear equations:
\begin{eqnarray}\label{xsceq16}
Z_{n \mathbf{k}}(i \omega_{j})&=&
1+\frac{T}{\omega_{j}N(\ef)}
\sum_{m j^{\prime}} \int\!\frac{\mathrm{d}\bq}{\Omega_{\rm BZ}} 
\frac{\omega_{j^{\prime}} Z_{m \mathbf{k}+\mathbf{q}}(i \omega_{j^{\prime}})}{\theta_{m \mathbf{k}+\mathbf{q}}(i \omega_{j^{\prime}})}\nonumber \\
&&\times \,
\lambda_{n \mathbf{k}, m \mathbf{k}+\mathbf{q}}(\omega_{j}-\omega_{j^{\prime}}) \nonumber\\
\end{eqnarray}
\begin{eqnarray}\label{xsceq17}
\chi_{n \mathbf{k}}(i \omega_{j})&=&
-\frac{T}{N(\ef)} 
\sum_{m j^{\prime}}\int\!\frac{\mathrm{d}\bq}{\Omega_{\rm BZ}} 
\frac{\varepsilon_{m \mathbf{k}+\mathbf{q}}-\cp+
\chi_{m \mathbf{k}+\mathbf{q}}(i \omega_{j^{\prime}})}{\theta_{m \mathbf{k}+\mathbf{q}}(i \omega_{j^{\prime}})}\nonumber \\
&&\times \,
\lambda_{n \mathbf{k}, m \mathbf{k}+\mathbf{q}}(\omega_{j}-\omega_{j^{\prime}}) \nonumber\\ 
\end{eqnarray}
\begin{eqnarray}
\phi_{n \mathbf{k}}(i \omega_{j}) &=&
\frac{T}{N(\ef)} 
\sum_{m j^{\prime}} \int\!\frac{\mathrm{d}\bq}{\Omega_{\rm BZ}} 
\frac{\phi_{m \mathbf{k}+\mathbf{q}}
(i \omega_{j^{\prime}})}{\theta_{m \mathbf{k}+\mathbf{q}}(i \omega_{j^{\prime}})} \nonumber \\
&&\times \,
\left[ \lambda_{n \mathbf{k}, m \mathbf{k}+\mathbf{q}}(\omega_{j}-\omega_{j^{\prime}})
-N(\ef)V_{n \mathbf{k}, m \mathbf{k}+\mathbf{q}}\right] \nonumber \\ \label{xsceq18}
\end{eqnarray}
This set of equations is supplemented with an equation for the electron number, which determines the chemical potential~$\cp$~\cite{marsiglio2008a,AGDbook1963}: 
\begin{equation}\label{xsceq19}
N_{\rm e}=\sum_{n} \int\!\frac{\mathrm{d}\bk}{\Omega_{\rm BZ}} \left[1-2 T \sum_{j}
\frac{\varepsilon_{n \mathbf{k}}-\cp + 
\chi_{n \mathbf{k}}(i\omega_j)}{\theta_{n \mathbf{k}}(i\omega_j)}\right],
\end{equation}
where $N_{\rm e}$ is the number of electrons per unit cell (see Supplementary Method~1 for a discussion about the electron number equation). 

Equations~(\ref{xsceq16})-(\ref{xsceq19}) involve electronic states that are not restricted to the Fermi surface or its immediate vicinity; hence, labeled as anisotropic full-bandwidth (FBW) Migdal-Eliashberg equations~\cite{Schrodi2019}.
We have recently implemented the above-described anisotropic FBW approach in the EPW code~\cite{EPW2}. 

The set of coupled equations can be solved self-consistently at various temperatures for the temperature-dependent superconducting gap \gap, given by:
\begin{equation}\label{xsceq20}
\Delta_{n \mathbf{k}}(i \omega_{j})=
\frac{\phi_{n \mathbf{k}}(i \omega_{j})}{Z_{n \mathbf{k}}(i \omega_{j})} .
\end{equation}
The Pad{\'{e}} approximation~\cite{vidberg1977a,leavens1985a} can then be used to obtain the continuation of $\Delta_{n \mathbf{k}}(i \omega_{j})$ from the imaginary to the real axis. The superconducting temperature \tc{} is the highest temperature at which $\phi_{n \mathbf{k}}(i \omega_{j})\neq 0$ has a nontrivial solution.

The contribution of the Coulomb interaction to the Eliashberg equation, through matrix elements $V_{n \mathbf{k}, m \mathbf{k}+\mathbf{q}}$, can be evaluated at the same level as the el-ph interaction for the simpler versions of the Eliashberg formalism~\cite{Sano2016,Sanna2018a,Errea2020-dz,Davydov2020a,Pellegrini2022}. To reduce the computational cost, however, the common approach is to replace the $N(\ef) V_{n \mathbf{k}, m \mathbf{k}+\mathbf{q}}$ terms with the semi-empirical Morel-Anderson pseudopotential~$\cs$~\cite{morel1962a}.
In practice, the numerical value of this parameter is connected with the cutoff frequency $\omega_{\rm max}$ of the Matsubara frequencies. With a typical choice of $\omega_{\rm max}$ being ten times the maximum phonon frequency $\omega_{\rm ph}$, a value of $\cs=0.1$--0.2 results in a satisfactory agreement with experiment for many applications. 
The $\cs$ can also be calculated from first-principles, using the double Fermi surface average of $V_{n \mathbf{k}, m \mathbf{k}+\mathbf{q}}$~\cite{Margine2016-bh,Heil2017,Heil2019,di2021}.

The numerical solution of the anisotropic FBW Migdal-Eliashberg equations is computationally very demanding. A common simplification of these equations consists in restricting the energy range close to the Fermi level by introducing the unity factor $\int_{-\infty}^{+\infty}\!\mathrm{d}\varepsilon \, \delta(\varepsilon_{n \mathbf{k}}-\varepsilon)$, and to assume that the DOS within this energy window is \mbox{constant~\cite{margine2013,allen1983b,choi2003a,marsiglio2008a,scalapino1966a,scalapino1969a,allen1976a,Marsiglio2020}}. It can be shown that, within these approximations, the energy shift $\chi_{n\bk}$ vanishes and the requirement in Eq.~(\eqref{xsceq19}) is automatically satisfied. As a result, only two equations for $Z_{n \mathbf{k}}(i \omega_{j})$ and $\phi_{n \mathbf{k}}(i \omega_{j})$ need to be solved self-consistently:
\begin{eqnarray}\label{xsceq21}
Z_{n \mathbf{k}}(i \omega_{j})&=&
1+\frac{\pi T}{N(\ef) \omega_{j}} \sum_{m 
j^{\prime}} \int\!\frac{\mathrm{d}\bq}{\Omega_{\rm BZ}} 
\frac{\omega_{j^{\prime}}}{\sqrt{\omega_{j^{\prime}}^{2}+
\Delta_{m \mathbf{k}+\mathbf{q}}^{2}(i \omega_{j^{\prime}})}} \nonumber \\
&&\times \, \lambda_{n \mathbf{k}, m \mathbf{k}+\mathbf{q}} 
(\omega_{j}-\omega_{j^{\prime}}) \delta(\varepsilon_{m \mathbf{k}+\mathbf{q}}-\ef)
\end{eqnarray}
\begin{eqnarray}\label{xsceq22}
Z_{n \mathbf{k}}(i \omega_{j}) \Delta_{n \mathbf{k}}(i \omega_{j}) &=&
\frac{\pi T}{N(\ef)}\sum_{m j^{\prime}}\int\!\frac{\mathrm{d}\bq}{\Omega_{\rm BZ}}\, 
\frac{\Delta_{m \mathbf{k}+\mathbf{q}}
(i \omega_{j^{\prime}})}{\sqrt{\omega_{j^{\prime}}^{2}+
\Delta_{m \mathbf{k}+\mathbf{q}}^{2}(i \omega_{j^{\prime}})}} \nonumber\\
&& \hspace{-2cm} \times \, \left[ \lambda_{n \mathbf{k}, m \mathbf{k}+\mathbf{q}} 
(\omega_{j}-\omega_{j^{\prime}})-N(\ef) V_{n \mathbf{k}, m \mathbf{k}+\mathbf{q}}\right] \delta
(\varepsilon_{m \mathbf{k}+\mathbf{q}}-\ef)\nonumber \\
\end{eqnarray}
Equations~\eqref{xsceq21} and \eqref{xsceq22} are the anisotropic FSR Migdal-Eliashberg equations. They have been the basis for the superconductivity calculations in the EPW code prior to the recent developments~\cite{margine2013,EPW}. 

\subsection{Sparse sampling of Matsubara frequencies}
\label{sec:theory_sparse}
All calculated quantities in the Migdal-Eliashberg equations depend on Matsubara frequencies, which are proportional to the absolute temperature. This results in a computational challenge since solving these equations at low temperatures (e.g., necessary for low-\tc{} superconductors) requires a larger number of frequencies within the same energy range.
As discussed in the previous section, for the FBW+$\mu$ method, one needs a Matsubara frequency cutoff 6-12 times the Fermi energy window in order to converge the chemical potential and, consequently, the superconducting gap energy, leading to a considerable increase in computational cost. While a frequency cutoff of 4\,eV has been found to be sufficient for calculations with FSR and FBW with fixed chemical potential at the Fermi level in the case of H$_3$S, adopting FBW+$\mu$ required a Matsubara frequency 2-3 times larger as shown in Supplementary Figure~1. 

The computational cost can be reduced by pruning the Matsubara frequencies \cite{Davydov2020a}. 
We implemented a sparse sampling scheme, described in the following, in the EPW code as an alternative to uniform sampling over the Matsubara frequencies.
Denoting with integer $N_{j}$ the numerical index for the $j^\mathrm{th}$ Matsubara frequency, $i\omega_{j}=i(2N_{j}+1)\pi T$, the uniform and sparse grids can be obtained as:
\begin{equation}\label{xsceq32}
N_{j}=j; \quad j=0,\pm 1, \pm 2,\pm 3, \cdots
\end{equation}
and
\begin{align}\label{xsceq33}
N_j=&N_{j-1}+\mathrm{INT}\left[ \exp\left(\frac{j}{N_\mathrm{max} W}\right)\right]\nonumber\\
&N_{0}=0 \text{ and } j=1, 2, 3, \cdots
\end{align}
where $\mathrm{INT[\ ]}$ is the rounding to the closest integer, $W$ is an adjustable weight factor, and $N_\mathrm{max}$ is the maximum Matsubara index
for a given energy cutoff $\omega_\mathrm{max}$ and temperature $T$:
\begin{equation}\label{xsceq34}
N_\mathrm{max} = 
\mathrm{INT}\left[ \frac{1}{2} \left( \frac{\omega_\mathrm{max}}{\pi T} - 1 \right) \right]
\end{equation}
Eq.~(\eqref{xsceq33}) can be used to generate a Matsubara frequency grid with indices $N_j \leq N_\mathrm{max}$. For negative indices, the corresponding frequency can be easily obtained as $\omega_{-(j+1)}=-\omega_{j}$ (with $j>0$). The resulting mesh is uniform at lower frequencies, contributing the most to the summation in the Migdal-Eliashberg equations, and becomes logarithmically sparser with increasing the Matsubara frequency index.

Increasing (decreasing) $W$ results in a denser (sparser) grid sampling. With the default setting of $W=1.0$, the sparse sampling scheme produces approximately $30\%$ fewer Matsubara frequencies than the uniform one, while the first $\sim$40$\%$ points of the grid are still uniformly distributed. Numerical tests show that this approach maintains the accuracy of more expensive calculations that use the full uniform grid.

We systematically evaluated this sampling approach by computing the superconducting gap function within the FBW+$\mu$ method using both the uniform and sparse sampling scheme over Matsubara frequencies up to 6\,eV for H$_3$S, D$_3$S, YH$_{6}$, and CaH$_{6}$. The primary outcomes are presented in Supplementary Figure~2, where the temperature-dependent behavior of \gap\ is plotted for FBW+$\mu$ (blue lines) and FBW+$\mu$+sparse (red lines). The results indicate that the sparse sampling method can accurately reproduce the superconducting gap structure obtained using the uniform sampling scheme. In summary, no significant difference was observed across all the compounds analyzed.
Among all the hydrides, the largest deviation was observed for D$_3$S, with a difference of only 3\,K compared to the uniform sampling, corresponding to a percentage difference of 1.8\,\%. For all other hydrides, the difference between the sparse and uniform schemes was only 1\,K. Thus, the sparse sampling can quantitatively reproduce the results obtained with the uniform sampling, but with a computational cost of approximately 40\,\% lower. 

\subsection{Computational details}
\label{sec:computational_details}
The electronic structure calculations are performed within the Kohn-Sham scheme~\cite{Kohn-Sham} of the density functional theory~\cite{DFT} as implemented in the Quantum ESPRESSO suite~\cite{QE1,QE2,QE3}. The exchange and correlation effects are treated within the Perdew-Burke-Ernzerhof parametrization~\cite{PBE} using scalar-relativistic optimized norm-conserving Vanderbilt pseudopotentials~\cite{ONCV1,ONCV2}. The Kohn-Sham orbitals are expanded in a plane-wave basis set with a kinetic-energy cutoff of 100\,Ry for H$_3$S, D$_3$S, and CaH$_6$, and 80\,Ry for YH$_6$ and BaSiH$_8$. The charge density is computed using $\Gamma$-centered Monkhorst-Pack \textbf{k}-meshes~\cite{k-mesh} of 24$^3$ \textbf{k}-points for H$_3$S, D$_3$S, and CaH$_6$, 16$^3$ \textbf{k}-points for YH$_6$, and 12$^3$ \textbf{k}-points for BaSiH$_8$. The Brillouin-zone integration employs a Methfessel-Paxton smearing~\cite{MP-smearing} of 0.01\,Ry for H$_3$S, D$_3$S, and BaSiH$_8$, and 0.04\,Ry for YH$_6$ and CaH$_6$. All lattice parameters and internal degrees of freedom were fully relaxed to reach a ground-state convergence of 10$^{-7}$\,Ry in the total energy and 10$^{-6}$\,Ry/$a_0$ for forces acting on the nuclei. BaSiH$_8$ is relaxed to a pressure of 30\,GPa, all other compounds to 200\,GPa.

The dynamical matrices and the linear variation of the self-consistent potential are calculated within the density-functional perturbation theory~\cite{DFPT} on a regular phonon grid of 4$^3$ \textbf{q}-points for H$_3$S, D$_3$S, and 6$^3$ \textbf{q}-points for YH$_6$, CaH$_6$, and BaSiH$_8$. The threshold for self-consistency is set to 10$^{-14}$ or lower.

The maximally localized Wannier functions (MLWFs) are constructed using the Wannier90 code~\cite{WANN1,WANN2}. In the case of H$_3$S and D$_3$S, 10 Wannier functions are used to describe the electronic states near the Fermi level. The Wannier orbitals are three H-$s$-like functions and seven functions with $s$, $p$, $d_{xy}$, $d_{xz}$, and $d_{yz}$ angular momentum states associated with the S site, with a spatial spread ranging from 0.77\,\AA$^2$ to 1.53\,\AA$^2$. For YH$_6$, six H-$s$-like projections and five Y-$d$-like functions are used to construct the initial guess, resulting in a spatial spread between 1.74\,\AA$^2$ and 1.86\,\AA$^2$. For CaH$_6$, besides the six H-$s$-like projections, we also use $sp$ and $sp^{3}$ hybrid orbital functions associated with the Ca site to construct the initial guess, yielding a spatial spread between 0.80\,\AA$^2$ and 1.20\,\AA$^2$. For BaSiH$_8$, we use H $s$, Si $s$, $p$, $d_{z^2}$, $d_{x^2-y^2}$, and Ba $p$, $d$ orbitals, giving a total of 22 Wannier functions, resulting in spatial spreads between \SI{1.04}{\angstrom^2} and \SI{1.79}{\angstrom^2}.

The fully anisotropic Migdal-Eliashberg equations~\cite{margine2013} are solved using the EPW code~\cite{EPW,EPW2}. Electron energies, phonon frequencies, and electron-phonon matrix elements are computed on fine grids containing 48$^3$ \textbf{k}- and \textbf{q}-points for H$_3$S, D$_3$S, YH$_6$, and CaH$_6$, and 30$^3$ \textbf{k}- and \textbf{q}-points for BaSiH$_8$. The lower boundary for the phonon frequency is set to 5\,cm$^{-1}$ for H$_3$S and D$_3$S, and 15\,cm$^{-1}$ for YH$_6$, CaH$_6$, and BaSiH$_8$. The width of the Fermi window is set to 2\,eV for H$_3$S, D$_3$S, YH$_6$, CaH$_6$, and BaSiH$_8$. We set the Matsubara frequency cutoff to $\omega_{\rm max}$ = 6\,eV. The smearing values for the energy-conserving $\delta$-function and for the sum over \textbf{q}-space in the el-ph coupling are set to 25\,meV and 0.05\,meV, respectively, for H$_3$S and D$_3$S, 150\,meV and 0.15\,meV for YH$_6$ and CaH$_6$, and \SI{100}{meV} and \SI{0.1}{meV} for BaSiH$_8$. We solved the equations adopting a Coulomb pseudopotential of $\cs = 0.16$ for all materials except BaSiH$_8$, where a value of 0.1 was chosen to be consistent with our previous works~\cite{lucrezi2022,lucrezi2022_quantum}. The continuation of the superconducting gap along the imaginary axis to the real energy axis is determined by applying the approximate analytic continuation using Padé functions~\cite{vidberg1977a,leavens1985a}.

The doping calculations within rigid band model are performed by shifting the Fermi level in EPW before interpolating the el-ph matrix elements. All anisotropic Migdal-Eliashberg calculations are performed for electronic energies of $\pm\SI{1}{eV}$ around the (shifted) Fermi level and a Matsubara cutoff of $\omega_{\rm max}=\SI{4}{eV}$, if not stated otherwise. 

\section{Data availability}
The authors confirm that the data supporting the findings of this study are available within the article. Further data and information is provided in the Supplemental Material, in the Zenodo database under accession code 10.5281/zenodo.10277399~\cite{zenodo}, and upon request.

\section{Code availability}
The code used to generate the data of this study is fully available within the open-source software package Quantum ESPRESSO v.7.2, which is distributed via the website quantum-espresso.org.

\section{Acknowledgements}
We thank Lilia Boeri and Warren Pickett for fruitful discussions and comments on the manuscript.

CH and RL acknowledge the Austrian Science Fund (FWF) Project No. P32144-N36. PPF gratefully acknowledges the S\~ao Paulo Research Foundation (FAPESP) under Grants 2020/08258-0 and 2021/13441-1. SH, HM, HP, and ERM acknowledge the support of the National Science Foundation under Grant No. DMR-2035518 and Grant No. OAC-2103991 for the development and interoperability of the superconductivity module of the EPW code. Calculations were performed on the Vienna Scientific Cluster (VSC5) and the Frontera supercomputer at the Texas Advanced Computing Center via the Leadership Resource Allocation (LRAC) award DMR22004 was used for code development and testing.

For the purpose of open access, the authors have applied a CC BY public copyright license to any author accepted manuscript version arising from this submission.

\section{Author Contributions}
RL and PPF contributed equally. RL and PPF performed the EPW calculations, prepared the figures and tables, and wrote the main draft. SH implemented the FBW formulation and the sparse sampling scheme in the superconductivity module of the EPW code, and HM and HP tested and expanded this module. 
CH and ERM supervised this project. 
All authors participated in the discussions and revised the manuscript.

\section{Competing interests}
The authors declare no competing interests.

\clearpage
\onecolumngrid
\printfigures

\clearpage
\printtables

\end{document}